\documentclass[aps,prb,showpacs,twocolumn,floats,epsfig,pdflatex]{revtex4}
\usepackage{amssymb}
\usepackage{amsbsy}
\usepackage{amsmath}
\usepackage{epsfig}
\usepackage{graphicx}
\begin{document}

\title {Phonon-induced topological insulation}

\author{Kush Saha}
\author{Ion Garate}
\affiliation{D\'epartement de Physique and Regroupement Qu\'eb\'ecois sur les Mat\'eriaux de Pointe, Universit\'e de Sherbrooke, Sherbrooke, Qu\'ebec, Canada J1K 2R1}
\date{\today}

\begin{abstract}
We develop an approximate theory of phonon-induced topological insulation in Dirac materials.
In the weak coupling regime, long wavelength phonons may favor topological phases in Dirac insulators with direct and narrow bandgaps.
This phenomenon originates from electron-phonon matrix elements, which change qualitatively under a band inversion. 
A similar mechanism applies to weak Coulomb interactions and spin-independent disorder; however, the influence of these on band topology is largely independent of temperature.
As applications of the theory, we evaluate the temperature-dependence of the critical thickness and the critical stoichiometric ratio for the topological transition in  CdTe/HgTe quantum wells and in BiTl(S$_{1-\delta}$Se$_{\delta})_2$, respectively.
\end{abstract}

\maketitle

\section{Introduction}
\label{intro}

Nearing a decade of rapid progress,\cite{fu} the field of topological insulators has become subject of textbooks.\cite{books}
Despite its incipient maturity, the subject remains active and broad with ongoing work ranging from exotic field theory\cite{exotic} to applied physics.\cite{yokoyama}
Above all, the mechanism of band inversion, responsible for the appearance of topological surface states, continues to captivate the imagination of theorists and experimentalists alike. 

At the interface between theory and experiment, there is an intense interest to identify tunable bandgap materials, where band inversion could be done and undone as a function of an experimentally controllable parameter.
Pressure,\cite{bahmy} electric fields, \cite{kim} compound stoichiometry,\cite{hasan} lasers\cite{diehl} and strong random alloying \cite{li} are some of the candidate agents that can switch topological phases on and off.
Recently, one of us has argued that temperature is an additional ``knob'' that may induce a band inversion.\cite{ion} 
This proposal originated from a study of electron-phonon interactions, whose effect in band topology had been previously overlooked.
The idea that phonons can change the band topology of an electronic system resonates with an upcoming line of research concerned with the effect of dissipative and thermal baths on topological materials.\cite{diss}

In Ref.~[\onlinecite{ion}], the origin for phonon-induced band inversion remained rather obscure.
The main objective of the present work is to provide a simple and intuitive understanding of the mechanism that underlies phonon-induced topological insulation.
This same mechanism applies to disorder- and Coulomb-interaction-induced topological insulation as well.
We begin (Sec.~\ref{bandgap}) by reviewing some fundamentals of bandgap renormalization in semiconductors.
This is a subject that has attracted steady attention in the last four decades;\cite{cardona_rev} nevertheless, the possibility that phonons could invert the bandgap of a Dirac material has not been contemplated. 
If one considers only the highest valence band and the lowest conduction band of a direct-gap insulator, then simple perturbation theory dictates that intraband (interband) electron-phonon scattering processes reduce (augment) the bandgap.
The total change in the bandgap is given by the sum of the competing interband and intraband parts.
In wide-gap semiconductors, interband transitions are suppressed and the bandgap decreases with temperature, regardless of the details of electron-phonon matrix elements.
However, in narrow-gap semiconductors, the competition between intraband and interband transitions becomes particularly sensitive to the matrix elements.
These matrix elements are peculiar in Dirac materials (Sec.~\ref{dirac}) because the momentum-space texture of the band eigenstates changes topology when the system undergoes a band inversion.
Due to this peculiarity, the dominant long-wavelength electron-phonon matrix elements are of intraband (interband) type in trivial (topological) Dirac insulators. 
Consequently,  the magnitude of the bandgap of a trivial Dirac insulator decreases with temperature, while it increases in a topological insulator.
This idea, which we elaborate in Sec.~\ref{phti}, is the main insight from the present work.
In Sec.~\ref{appl}, we apply the theory to HgTe/CdTe quantum wells and BiTl(S$_{1-\delta}$Se$_{\delta})_2$.
Because the renormalized Dirac mass and the renormalized bandgap differ from one another, the appearance of topological surface states does not occur simultaneously with a band inversion.
This may help explain the ``topological proximity effect'' observed in BiTl(S$_{1-\delta}$Se$_{\delta})_2$.
In Sec.~\ref{summ}, we collect the main ideas and discuss their relevance in materials with complex electronic and phononic structures.

\section{Bandgap renormalization}
\label{bandgap}

We begin by reviewing the phonon-induced renormalization of the bandgap in semiconductors.\cite{cardona_rev}
The Hamiltonian of a perfectly periodic crystal is 
\begin{equation}
{\cal H}_0=\sum_{\bf k}\sum_{\sigma\sigma'}\sum_{\tau\tau'} \langle \sigma \tau | h_{\bf k}| \sigma'\tau'\rangle c^\dagger_{{\bf k}\sigma\tau} c_{{\bf k}\sigma'\tau'},
\end{equation} 
where ${\bf k}$ is the crystal momentum, $h_{\bf k}$ is the Bloch Hamiltonian and $c_{{\bf k}\sigma\tau}$ is an operator that annihilates an electron with momentum ${\bf k}$, spin quantum number $\sigma$ and orbital quantum number $\tau$.
The Bloch Hamiltonian has eigenvalues $E_{{\bf k}n}$ and eigenstates $|\psi_{{\bf k} n}\rangle=\exp(i {\bf k}\cdot{\bf r}) |{\bf k} n\rangle/\sqrt{V}$, where $V$ is the crystal volume, $n$ is the Bloch band label and $|{\bf k} n\rangle$ is a multicomponent spinor whose spatial dependence has the periodicity of the lattice.

Small deviations of the ions from their equilibrium positions couple to the electron density and result in an electron-phonon interaction ${\cal V}\simeq {\cal V}^{(1)}+{\cal V}^{(2)}$, where
\begin{align}
\label{eph}
{\cal V}^{(1)} &= \int d{\bf r}\, \rho({\bf r})\sum_{j} {\bf Q}_j\cdot {\boldsymbol\nabla} V_{\rm ei}({\bf r}-{\bf R}_j^{0})\nonumber\\
{\cal V}^{(2)} &={1 \over 2}\int d{\bf r}\, \rho({\bf r}) \sum_{j} ({\bf Q}_{j}\cdot{\boldsymbol\nabla})^2 V_{\rm ei}({\bf r}-{\bf R}_{j}^0).
\end{align}
Here, $R_j^{0}$ is the equilibrium position of the $j$-th ion, ${\bf r}$ is the electron coordinate, ${\boldsymbol\nabla}=\partial/\partial{\bf r}$, $V_{\rm ei}({\bf r})$ is the electron-ion potential,
\begin{equation}
{\bf Q}_j=i\sum_{\bf k} e^{i {\bf k}\cdot{\bf R}^{(0)}_j} \left(\frac{\hbar}{2\rho_A V \omega_{\bf k}}\right)^{1/2}{\bf e}_{\bf k}\, (a_{\bf k}+a^\dagger_{-\bf k}) 
\end{equation}
is the ionic displacement from equilibrium (assumed to be small), $a_{\bf k}$ is an operator that destroys a phonon with momentum ${\bf k}$, $\omega_{\bf k}$ is the phonon frequency,  $\rho_A$ is the atomic density and ${\bf e}_{\bf k}$ is the polarization vector of the phonon mode.
In addition, the electron density operator is
\begin{equation}
\label{rho}
\rho({\bf r})=\frac{1}{V}\sum_{\bf q} e^{-i {\bf q}\cdot {\bf r}}\rho_{\bf q}\,\,\,;\,\,\,\rho_{\bf q}=\sum_{{\bf k}\sigma\tau}  c^\dagger_{{\bf k}\sigma\tau} c_{{\bf k-{\bf q}}\sigma\tau}
\end{equation}
in the plane wave basis.
For simplicity, we have considered one atom per lattice site and, for brevity, we have omitted the index that labels different phonon modes.
Furthermore, Eq.~(\ref{eph}) is local in real space and thus does not capture phonon-induced changes in the electronic hopping amplitude; the effect of these terms will be briefly discussed in Sec.~\ref{summ}.

Equations~(\ref{eph}) and (\ref{rho}) together evidence that the electron-phonon interaction conserves spin and orbital quantum numbers in the plane wave basis.
This fact will figure prominently in the mechanism for phonon-induced topological insulation (cf. Sec.~\ref{phti}).
At any rate, since the the electronic eigenstates do not generally have well-defined spin and orbital quantum numbers, phonons do scatter electrons between different bands.
This becomes apparent by rewriting Eq.~(\ref{rho}) in the band eigenstate basis,
\begin{equation}
\label{rho1}
\rho_{\bf q}=\sum_{{\bf k} {\bf k'} n n'}\sum_{\bf G} \delta_{{\bf k'}+{\bf q}-{\bf k},{\bf G}}\langle {\bf k} n|{e^{i {\bf G}\cdot{\bf r}}|\bf k'}\, n'\rangle c^\dagger_{{\bf k} n} c_{{\bf k'} n'},
\end{equation}
where $c_{{\bf k}n}$ annihilates an electron with momentum ${\bf k}$ in band $n$, ${\bf G}$ is a reciprocal lattice vector and 
\begin{equation}
\label{matel}
\langle  {\bf k} n|e^{i{\bf G}\cdot{\bf r}}| {\bf k'} n'\rangle\equiv\int_{\rm cell} d{\bf r} e^{i {\bf G}\cdot{\bf r}} u_{{\bf k} n}^* ({\bf r}) u_{{\bf k'}n'}({\bf r}).
\end{equation}
Here, $u_{{\bf k} n}({\bf r})\equiv\langle {\bf r}|{\bf k} n\rangle$ and the spatial integration is over the unit cell.
If the Bloch eigenstates were plane waves (which would approximately be the case in simple metals), Eq.~(\ref{matel}) would be nonzero only for $G=0$.
In Dirac insulators, the Bloch states are not plane waves and the $G\neq 0$ terms (Umklapp processes) do not vanish.
Yet, hereafter we neglect Umklapp processes on the basis that (i) we consider the coupling of electrons to long wavelength phonons (deformation potential coupling), and (ii) we model the electronic structure with ${\bf k}\cdot{\bf p}$ Hamiltonians that are tailored to small momenta in the vicinity of the bandgap minimum.
Under this proviso, Eq.~(\ref{eph}) can be rewritten as\cite{mahan}
\begin{align}
\label{vv}
{\cal V}^{(1)}&=\sum_{\bf q} g_{\bf q} \rho_{\bf q}(a^\dagger_{-\bf q}+a _{\bf q})\nonumber\\
{\cal V}^{(2)}&=\sum_{{\bf k},{\bf q}} \lambda_{{\bf k}{\bf q}} \rho_{\bf q} (a_{\bf k}+a^\dagger_{-{\bf k}})(a_{-{\bf k+q}}+a^\dagger_{\bf k-q}),
\end{align}
where the expressions for $g_{\bf q}$ and $\lambda_{{\bf k}{\bf q}}$ are listed in Appendix A.
Both ${\cal V}^{(1)}$ and ${\cal V}^{(2)}$ modify the electronic band structure.
At zero temperature, the renormalized energy levels are 
\begin{equation}
E^*_{{\bf k} n } \simeq E_{{{\bf k}n}}+\sum_{n'{\bf q}}|g_{\bf q}|^2 {|\langle {\bf k} n|{\bf k-q} n'\rangle|^2\over E_{{\bf k} n}-E_{{\bf k-q} n'}},
\label{eqne}
\end{equation}  
where we have neglected the frequency of phonon modes in the denominator.
This is a good approximation for the purposes of the present work (cf. Sec.~\ref{phti}).
In addition, in the derivation of Eq.~(\ref{eqne}) we have used (cf. App.~\ref{ap_dw})
\begin{equation}
\label{dw}
\langle {\bf k} n; 0|{\cal V}^{(2)}|{\bf k} n'; 0\rangle=0,
\end{equation}
where $|{\bf k} n; N\rangle\equiv  c^\dagger_{{\bf k}n}|FS\rangle\otimes|N\rangle$, $|FS\rangle$ is the Fermi sea and $|N\rangle$ is a Fock state of $N$ phonons.
The matrix element in Eq.~(\ref{dw}) is known as the Debye-Waller term.
When short-wavelength phonons are included, the Debye-Waller term is nonzero and contributes to the renormalization of energy levels at the same order as Eq.~(\ref{eqne}).
In such case, the electron-phonon matrix elements are usually computed from first-principles pseudopotential methods,\cite{cohen} which show that the bandgap renormalization of common semiconductors often contains a significant Debye-Waller component.
The possible impact of the Debye-Waller term on band topology will be briefly discussed in Sec.~\ref{summ}.

\begin{figure}
\rotatebox{0}{\includegraphics*[width=\linewidth]{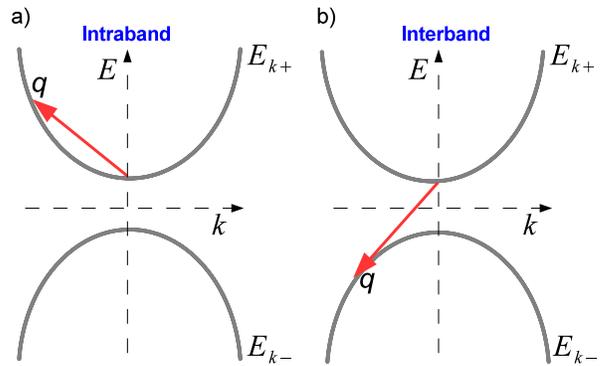}}
\caption{(Color online) Schematic diagram for (a) intraband and (b) interband electronic transitions induced by phonons in a two-band insulator with a direct gap at the zone center (each band may be degenerate).
It is assumed that additional conduction and valence bands are sufficiently far in energy so that they do not  contribute substantially to bandgap renormalization.
Such bandgap renormalization involves the evaluation of Eq.~(\ref{eqne}) for ${\bf k}={\bf 0}$.
Accordingly, the relevant electron-phonon scattering matrix elements connect ${\bf k}=0$ with ${\bf k}={\bf q}$, where ${\bf q}$ is the phonon momentum. 
Since $E_{{\bf 0} +}\leq E_{{\bf k}+}$ and $E_{{\bf 0}-}\geq E_{{\bf k}-}$ for any ${\bf k}$, intraband transitions lead to a decrease of the bandgap ($E_{{\bf 0}+}-E_{{\bf 0}-}$).
A similar argument reveals that interband transitions increase the bandgap. 
} \label{fig1}
\end{figure}

In Eq.~(\ref{eqne}), it is instructive to separate the sum over intermediate states onto interband ($n'\neq n$) and intraband ($n'=n$) parts.
For a direct gap semiconductor, a glance at the energy denominators of Eq.~(\ref{eqne}) and Fig.~\ref{fig1} reveals that intraband transitions decrease the bandgap at $k=0$, while interband transitions increase it.
In this paper, ``interband transition'' refers to a transition that takes place between the valence band and the conduction band edges.
Transitions between energy-degenerate bands are counted as ``intraband''.
There are, of course, interband transitions between different (non-degenerate) valence bands as well as between different conduction bands.
These interband transitions could {\em a priori} lead to a decrease of the bandgap.
However, for Dirac insulators in the vicinity of a topological phase transition, there is often a large energy separation between the lowest conduction band and the rest of conduction bands.
Likewise, the highest valence band is typically well separated in energy from the rest of the valence bands. 
Under these conditions, the leading interband contribution emerges from transitions between the highest valence band and the lowest conduction band, the rest being suppressed by relatively large energy denominators in Eq.~(\ref{eqne}).
In sum, the net change in the bandgap depends on the relative strength of the intraband and interband contributions. 

For wide-gap semiconductors, the interband contribution is depleted by a large energy denominator in Eq.~(\ref{eqne}); therefore, electron-phonon interactions decrease the bandgap.
However, for narrow-gap semiconductors, the energy denominators are no longer large enough to rule out the interband part.
Instead, the competition between intraband and interband contributions depends sensitively on the magnitude of the $G=0$ electron-phonon matrix elements, $ g_{\bf k-k'}\langle {\bf k} n| {\bf k}' n'\rangle$.
As we explain below (cf. Sec.~\ref{phti}), the key for the phonon-induced topological insulation is that intraband matrix elements dominate on the trivial side of a topological phase transition, whereas interband matrix elements take over on the nontrivial side of said phase transition.
In other words, phonons decrease the bandgap when the Dirac insulator is trivial,  while they increase the bandgap when the Dirac insulator is topological.
Albeit unusual, this feature seems generic to Dirac insulators with deformation potential coupling to phonons because it arises due to the change in the eigenstates' momentum-space texture across a topological phase transition.

\section{Dirac mass renormalization}
\label{dirac}

The aim of this paper is to extract qualitative insights from low-energy effective models, rather than from more accurate but less transparent first principles calculations. 
The minimal ${\bf k}\cdot{\bf p}$ Hamiltonian that captures the low-energy properties of a time-reversal- and inversion-symetric 2D or 3D Dirac insulator with a bandgap minimum at the Brillouin zone center \cite{franz} is 
\begin{equation}
\label{hD}
h_{\bf k}=d_{0,{\bf k}}+ {\bf d}_{\bf k}\cdot{\boldsymbol\sigma}\tau^x+M_{\bf k}\tau^z,
\end{equation}
where $\sigma^i$ and $\tau^i$ are Pauli matrices in spin and orbital space (respectively), $d_{0,{\bf k}}=\gamma k^2 a^2$, 
$d_{i,{\bf k}}=-\alpha k_i a$ ($i\in\{x,y\}$ in 2D and $i\in\{x,y,z\}$ in 3D), $M_{\bf k}=m+\beta k^2 a^2$, 
$a$ is the lattice constant that acts as an ultraviolet cutoff,  and $(\gamma,\beta,\alpha, m)$ are material parameters with units of energy.
In particular, $m$ is the Dirac mass of the quasiparticles at $k=0$.
The parameter $\alpha$ determines the velocity of Dirac quasi-particles, whereas $\gamma$ models the particle-hole asymmetry of the band structure. 
A lattice regularization of Eq.~(\ref{hD}) may be introduced in the usual way, but it does not affect the main results substantially.

The eigenvalues of Eq.~(\ref{hD}) are a pair of doubly degenerate conduction and valence bands with energies 
\begin{equation}
E_{{\bf k}\pm}=d_{0,{\bf k}}\pm\epsilon_{\bf k}\,\,\,  ; \,\,\,\epsilon_{\bf k}\equiv \sqrt{{\bf d}_{\bf k}\cdot{\bf d}_{\bf k}+M_{\bf k}^2}.
\label{engy}
\end{equation}
It is straightforward to obtain the corresponding eigenvectors analytically; these will be used to derive the results below. 
We assume that the bandgap at $k=0$ is parametrically smaller than the gap at any other time-reversal-invariant momentum (TRIM), i.e. we assume $|\beta|\gg|m|$.
This scenario comprises most real Dirac insulators.
The band topology of the Dirac insulator is then determined by the sign of $m \,\beta$: if positive (negative), the insulator is trivial (topological).
From here on we take $\beta>0$ without loss of generality, and thus $m>0$ ($m<0$) describes a trivial (topological) insulator.
Although it plays no role in determining the band topology of noninteracting Dirac insulators,  $\gamma$ can alter the band topology under the presence of electron-phonon interactions.

Phonons affect the band topology of a Dirac insulator by renormalizing the Dirac mass, $m\to m^*$.
In particular, if $m$ and $m^*$ have the opposite sign, phonons induce a topological phase transition.\cite{ion}
For real Dirac insulators, this is susceptible to occur only at the TRIM where the bandgap is smallest (at $k=0$ in our model) because the Dirac masses at all other TRIM are typically large in magnitude compared to the energy-scale of the electron-phonon interaction.
The renormalization of the Dirac mass due to electron-phonon interactions can be obtained from the self-energy,
\begin{align}
\label{se}
&\Sigma_{n n'}({\bf k},i\omega) =\sum_{{\bf q} n''} g_{\bf q}^2 \langle{\bf k} n|{\bf k}-{\bf q}\, n''\rangle\langle {\bf k-q} n''| {\bf k} n'\rangle\nonumber\\
&\times\left[\frac{1+n_{B\bf q}-f_{{\bf k-q} n''}}{i\omega-\xi_{{\bf k-q} n''}-\omega_{\bf q}}+\frac{n_{B\bf q}+f_{{\bf k-q} n''}}
{i\omega-\xi_{{\bf k-q} n''}+\omega_{\bf q}}\right],
\end{align}
where $n,n'\in\{1,2,3,4\}$ are band labels, $\xi_{{\bf k}n}=E_{{\bf k}n}-\epsilon_F$ with $E_{{\bf k}1}=E_{{\bf k}2}=E_{{\bf k}+}$ and $E_{{\bf k}3}=E_{{\bf k}4}=E_{{\bf k}-}$; $\epsilon_F$ is the Fermi energy, 
$n_{B\bf q}=[\exp(\omega_{\bf q}/T)-1]^{-1}$ is the phonon occupation number, $f_{{\bf k}n}=[\exp(\xi_{{\bf k}n}/T)+1]^{-1}$
 is the fermion occupation number and $\omega= (2 l+1)\pi T$ ($l\in\mathbb{Z}$) is the fermionic Matsubara frequency.

Hereafter we concentrate on the ${\bf k=0}$ self-energy,
\begin{equation}
\label{se0}
\Sigma_{n n'} ({\bf 0},i\omega)=\Sigma_{0}({\bf 0},i\omega)\delta_{n n'}+\Sigma_z ({\bf 0},i\omega)\tau^z_{n n}\delta_{n n'},
\end{equation}
where $\Sigma_0$ and $\Sigma_z$ are related to the renormalization of the Fermi energy and the Dirac mass, respectively (cf. Eq.~(\ref{hD})). 
The self-energy at $k=0$ is diagonal and doubly degenerate due to the combined time-reversal and inversion symmetry of Eq.~(\ref{hD}). 
After simple algebra, we arrive at
\begin{align}
\label{sigma}
&\Sigma_0 ({\bf 0},i\omega) \simeq\sum_{\bf q} g_{{\bf q},\rm eff}^2(T)\frac{i\omega+\epsilon_F-d_{0,{\bf q}}}{(i\omega+\epsilon_F-d_{0,{\bf q}})^2-\epsilon_{\bf q}^2}\nonumber\\
&\Sigma_z ({\bf 0},i\omega) \simeq\sum_{\bf q} g_{{\bf q},\rm eff}^2(T)\frac{M_{\bf q}}{(i\omega+\epsilon_F-d_{0,{\bf q}})^2-\epsilon_{\bf q}^2},
\end{align}
where we have neglected the phonon frequency in the denominators and have defined
\begin{equation}
\label{geff}
g_{{\bf q},\rm eff}^2(T)\equiv g_{\bf q}^2 (1+2 n_{B{\bf q}}).
\end{equation}
From Eq.~(\ref{se0}), the zero-temperature renormalized Dirac mass can be read off as\cite{wang} 
\begin{equation}
\label{mst}
m^*=m+\Sigma_z({\bf 0},0).
\end{equation}
A natural generalization to finite temperature is 
\begin{equation}
\label{mst2}
m^*(T)=m+{\rm Re}\,[\Sigma_z({\bf 0},i \pi T)],
\end{equation}
where we have recognized that the lowest Matsubara frequency is $\pm \pi T$.
When temperature is low compared to the bandwidth of the electronic bands (which is in fact the case of interest), ${\rm Re}\,[\Sigma_z({\bf 0},i \pi T)]\simeq \Sigma_z({\bf 0},0)$ and the entire temperature-dependence of $m^*(T)$ originates from the phonon occupation factor in Eq.~(\ref{geff}).
The physical consequences of a temperature-dependent Dirac mass will be discussed in Sec.~\ref{appl}.

An important aspect of Eq.~(\ref{sigma}) is that $\Sigma_z ({\bf 0},0)$ is largely independent of $m$ when $|m|$ is small.
Whether phonons favor a trivial or a topological phase is thus independent of whether the bare Dirac insulator is trivial or topological (insofar as the gap is small).
The physical reason behind this result will be described in Sec.~\ref{phti}.

Equations~(\ref{se0}) and (\ref{sigma}) are formally very similar to the ones that appear in the theory of the topological Anderson insulator.\cite{li}
The main difference arises in the temperature-dependence of the self-energy, which is negligible for static disorder and significant for phonons.
Temperature may be regarded as a knob to effectively tune the strength of electron-phonon coupling (cf. Eq.~(\ref{geff})).
In fact, the experimental fingerprint for the phonon-induced bandgap renormalization in semiconductors is its temperature-dependence. 

We conclude this section by discussing the relation between the renormalized Dirac mass and the renormalized bandgap.\cite{wang}
For small phonon frequency and at zero temperature, $\Sigma_{nn}({\bf k},\xi_{{\bf k}n})$ agrees with Eq.~(\ref{eqne}).
Accordingly, the renormalized bandgap at ${\bf k}={\bf 0}$ is $E_g^*\equiv E_c^*-E_v^*$, where 
\begin{align}
\label{rengap}
E_c^* &\simeq  m+\Sigma_{0}({\bf 0},E_c^*-\epsilon_F)+\Sigma_z({\bf 0}, E_c^*-\epsilon_F)\nonumber\\
E_v^* &\simeq  -m+\Sigma_0({\bf 0}, E_v^*-\epsilon_F)-\Sigma_z({\bf 0}, E_v^*-\epsilon_F).
\end{align}
Note that $E_g^*$ depends on temperature and may be either positive (normal band ordering) or negative (inverted band ordering).
Due to the frequency-dependence of the electron-phonon self-energy, $E_g^*\neq 2 m^*$.
Conceptually, the difference between the renormalized bandgap (which is the bulk gap measured e.g. in ARPES) and the renormalized Dirac mass (which dictates the existence of topological surface states) implies that the emergence or disappearance of topological surface states does not go hand in hand with the closing of the bulk quasiparticle gap.
In Sec.~\ref{appl}B, we argue that this difference may help explain recent ARPES measurements\cite{hasan1} that have probed the vicinity of the topological phase transition in BiTl(S$_{1-\delta}$ Se$_\delta$)$_2$.

\section{Phonon-induced topological insulation}
\label{phti}

Having reviewed the preliminary concepts, we are ready to discuss when and why phonons favor topological phases in Dirac insulators. 
The objective of this section is to extract some general principles that govern the fate of band topology in presence of electron-phonon interactions.
These principles should be valid beyond the toy model that we use to illustrate them.

Starting from Eq.~(\ref{hD}), we denote the self-energy for the positive-energy and negative-energy bands (each of which is doubly degenerate)  as $\Sigma_+$ and $\Sigma_-$, respectively.
In the spirit of Sec.~\ref{bandgap}, we separate the intraband and interband contributions:
\begin{equation}
\Sigma_\pm({\bf 0},i\omega)=\Sigma^{\rm intra}_\pm({\bf 0},i\omega)+\Sigma_\pm^{\rm inter}({\bf 0},i\omega).
\label{eqnse}
\end{equation}
From Eq.~(\ref{se}), it follows that 
\begin{align}
\label{intraer}
\Sigma^{\rm intra}_\pm({\bf 0},i\omega) &=\sum_{\bf q} g_{\bf q,{\rm eff}}^2 \,\frac{|V^{\rm intra}_{\bf q}|^2}{i\omega+\epsilon_F-E_{{\bf q}\pm}}\nonumber\\
\Sigma^{\rm inter}_\pm({\bf 0},i\omega) &=\sum_{\bf q} g_{\bf q,{\rm eff}}^2 \,\frac{|V^{\rm inter}_{\bf q}|^2}{i\omega+\epsilon_F-E_{{\bf q}\mp}},
\end{align}
where the intraband and interband matrix elements ($V^{\rm intra}_{\bf q}$ and $V^{\rm inter}_{\bf q}$) describe the probability amplitude for electronic transitions from ${\bf k}={\bf 0}$ to ${\bf k}={\bf q}$.
For intraband transitions, the initial and final scattering states are in bands with the same energy dispersion; for interband transitions, one of them is in the conduction band and the other one is in the valence band.
The explicit expressions for the matrix elements are
\begin{align}
\label{me1}
|V^{\rm intra}_{\bf q}|^2 &=\frac{1}{2}\left(1+{\rm sgn}(m)\frac{M_{\bf q}}{\epsilon_{\bf q}}\right)\nonumber\\
|V^{\rm inter}_{\bf q}|^2 &=\frac{1}{2}\left(1-{\rm sgn}(m)\frac{M_{\bf q}}{\epsilon_{\bf q}}\right),
\end{align}
where $M_{\bf q}$ and $\epsilon_{\bf q}$ were defined below Eq.~(\ref{hD}) and in Eq.~(\ref{engy}), respectively.
Note that $|V^{\rm intra}_{\bf q}|^2, |V^{\rm inter}_{\bf q}|^2\in[0,1]$ because they correspond to transition probabilities.
In order to interpret Eq.~(\ref{me1}), it is useful to recall that $\tau^z$ is a good quantum number at ${\bf k}={\bf 0}$ (cf. Eq.~\ref{hD}) and that phonons conserve the orbital pseudospin (cf. Eqs.~(\ref{rho}) and ~(\ref{vv})).
Then, Eq.~(\ref{me1}) gives a measure of how parallel the orbital pseudospin at ${\bf k}={\bf q}$ is with respect to that at ${\bf k}={\bf 0}$.
This is illustrated in Figs.~\ref{mel}a and \ref{mel}b.
The reason why Eq.~(\ref{me1}) depends on the sign of the bare Dirac mass $m$ is because the orbital pseudospin at ${\bf k}={\bf 0}$ flips direction when $m$ changes sign.
This, in turn, is the essence of band inversion.

Combining Eqs.~(\ref{sigma}), (\ref{eqnse}) and (\ref{intraer}), the renormalized Dirac mass is obtained from
\begin{equation}
\label{sz}
\Sigma_z ={\rm sgn}(m)(\Sigma_+-\Sigma_-)/2, 
\end{equation}
where the momentum and frequency arguments are ${\bf 0}$ and $\pi T$, respectively. 
The sign change in the expression for $\Sigma_z$ between $m>0$ and $m<0$ is due to the band inversion.
When $m=0$, the labeling of ``intraband'' and ``interband'' becomes ambiguous because the conduction and valence bands touch at $k=0$.
In this case one may take either ${\rm sgn}(m)=1$ or ${\rm sgn}(m)=-1$ in Eqs.~({\ref{me1}) and~(\ref{sz}), and one arrives at the same expression for $m^*$.

\begin{figure}
\rotatebox{0}{\includegraphics*[width=\linewidth]{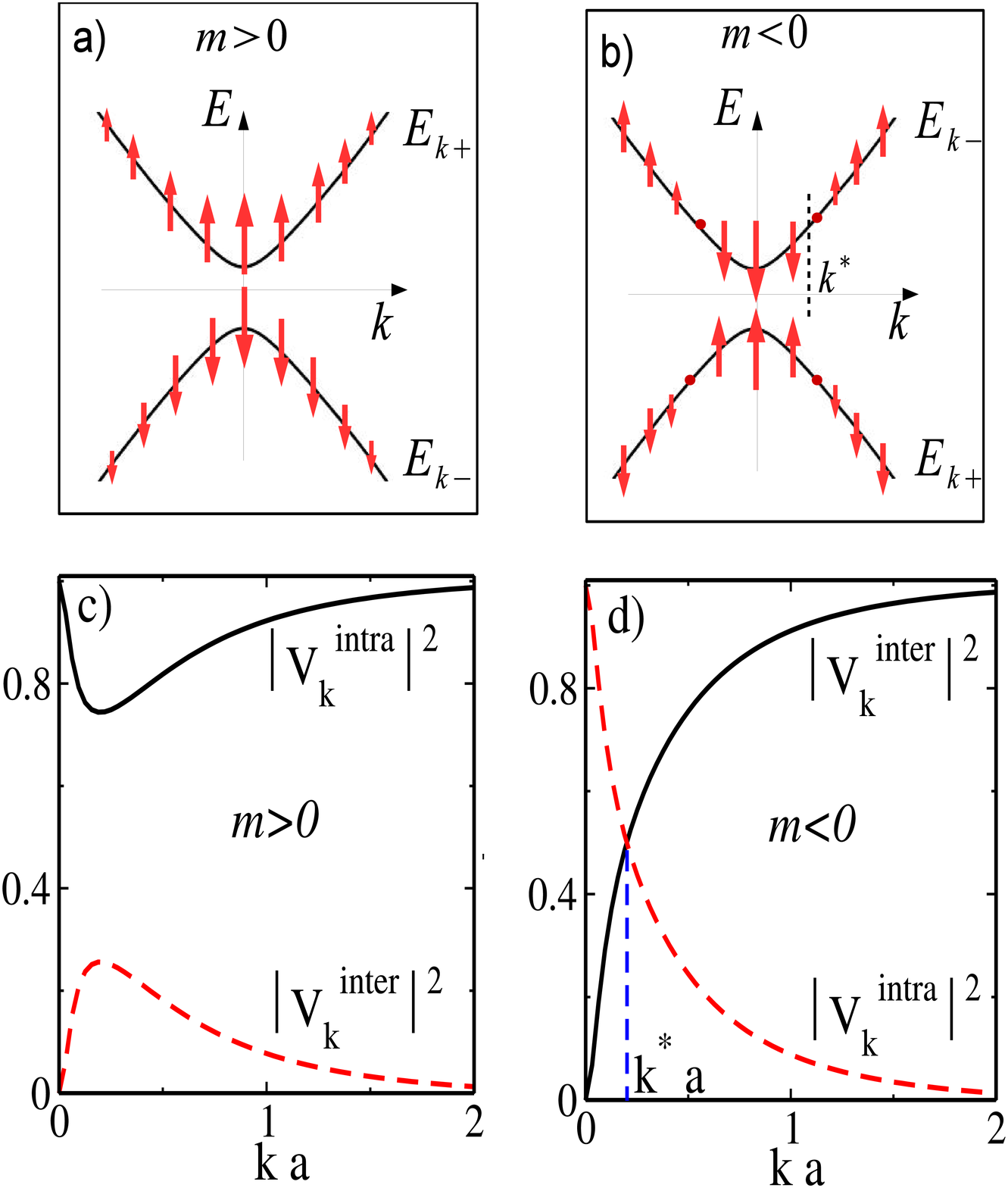}}
\caption{(Color online) (a) and (b): Expectation value of ${\boldsymbol\tau}$ (arrows) as a function of momentum in (a) a trivial insulator  and in (b) a topological insulator.
From the analytical expressions for the band eigenstates, we obtain $\langle {\boldsymbol\tau}\rangle_{{\bf k}\pm}=\hat{z}M_{\bf k}/E_{{\bf k}\pm}$. 
Degenerate energy bands give the same contribution to $\langle\tau^z\rangle$; in contrast, $\langle\tau^x\rangle_{{\bf k}\pm}=\langle\tau^y\rangle_{{\bf k}\pm}=0$ upon summing over each pair of degenerate bands. 
(c) and (d):  Electronic scattering probability from the zone center to a state with momentum ${\bf k}$. 
The scattering probability is maximized if the spin and pseudospin of the state at momentum ${\bf k}$ are parallel to those of the state at the zone center.  
(c): In a trivial insulator, intraband matrix elements dominate at all momenta transfer because $\langle\tau^z\rangle$ does not change sign with ${\bf k}$.
(d): In a topological insulator, intraband matrix elements dominate for $k<k^*$, where $k^*$ is defined via $M_{\bf k^*}=0$, while interband matrix elements dominate for $k>k^*$.}
\label{mel}
\end{figure}

Let us first consider $m>0$, which corresponds to a trivial bare Dirac insulator.
In this case,  $M_{\bf q}>0$ and thus $|V^{\rm intra}_{\bf q}|>|V^{\rm inter}_{\bf q}|$ for any ${\bf q}$.
Namely, phonons tend to scatter the ${\bf k}={\bf 0}$ electron into another state in the same (or different-but-degenerate) band because the orbital pseudospin is more aligned therein (cf. Fig.~\ref{mel}a).
The importance of this point becomes easier to grasp if we consider an undoped ($\epsilon_F=0$) and particle-hole symmetric ($d_{0,{\bf q}}=0$) insulator at zero temperature. 
In this case $\Sigma_+^{\rm intra}({\bf 0},0)<0<\Sigma_+^{\rm inter}({\bf 0},0)$, with $|\Sigma_+^{\rm intra}({\bf 0},0)|>\Sigma_+^{\rm inter}({\bf 0},0)$.
Thus $\Sigma_+({\bf 0},0)<0$ and, by particle-hole symmetry, $\Sigma_-({\bf 0},0)=-\Sigma_+({\bf 0},0)$.
Accordingly, Eqs.~(\ref{mst}) and~(\ref{sz}) dictate that $m^*<m$.
In other words, phonons drive an undoped and particle-hole symmetric trivial Dirac insulator towards the topological phase, solely because the intraband matrix elements prevail over the interband matrix elements.
For a sufficiently strong electron-phonon coupling (or sufficiently small $m$), $m^*<0$ and phonons induce a topological insulating phase in an otherwise trivial insulator.\cite{scba}
At the same time, phonons decrease the renormalized gap and may as well invert it, although $E_g^*$ does not change sign at the same time as $m^*$.

\begin{figure}
\includegraphics[scale=0.3]{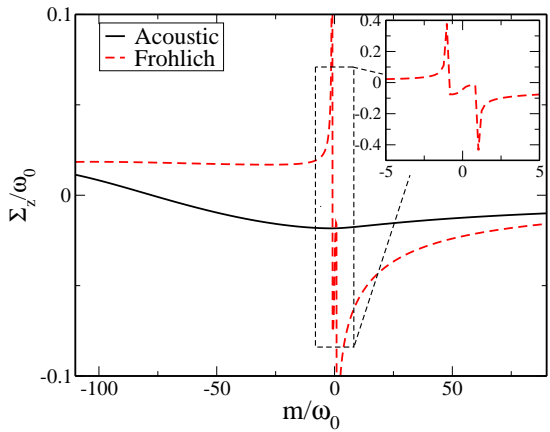}
\caption{(Color online) Dependence of the electron-phonon self-energy on the bare Dirac mass for different types of electron-phonon couplings $g_{\bf q}$. 
The axes are normalized in units of $\omega_0=20 {\rm meV}$, the optical phonon frequency used for the Fr\"ohlich coupling. 
The component of the self-energy that is responsible for renormalizing the Dirac mass, $\Sigma_z$, changes sign as a function of $m$ when $m$ becomes increasingly negative.
The reason for this is explained in the main text.
The value of $|m|$ for which the sign change occurs depends on the particular type of electron-phonon coupling $g_{\bf q}$.
If $g_{\bf q\to 0}$ is finite (or zero), then the sign change occurs when $q^*\sim a^{-1}$. 
In contrast, if $g_{\bf q\to 0}$ diverges (which is the case for Fr\"ohlich coupling in absence of screening), then small momenta transitions receive higher weight and $\Sigma_z$ changes sign at a considerably smaller value of $q^*$ (see inset).
}
\label{dif_phn}
\end{figure}

Next, we consider the case $m<0$. 
Here, $M_{\bf q}$ changes from negative to positive as $q$ varies from $0$ to $\pi/a$.
The sign change, which occurs at $q\, a \simeq  q^*\, a \equiv\sqrt{|m|/\beta}$, reflects a nontrivial texture of the orbital pseudospin in momentum space (cf. Fig.~\ref{mel}b).
Since phonons favor transitions between aligned orbital pseudospins, it follows that $|V^{\rm intra}_{\bf q}|>|V^{\rm inter}_{\bf q}|$ for $q<q^*$ and $|V^{\rm intra}_{\bf q}|<|V^{\rm inter}_{\bf q}|$ for $q>q^*$. 
Let us once again consider an undoped Dirac insulator with particle-hole symmetry.
If the main contribution to the self-energy originated from electron-phonon scattering processes with small momentum transfer ($q <q^*$),  
the intraband contribution would dominate, thereby resulting in $m^*>m$ and $E_g^*>E_g$ (i.e. $|m^*|<|m|$ and $|E_g^*|<|E_g|$).
In this scenario, phonons would have driven a topological insulator towards the trivial phase and, in conjunction with the preceding paragraph, we would have concluded that phonons favor the Dirac semimetal phase irrespective of whether the insulator was topological or trivial to begin with.
However, in a narrow gap Dirac insulator, $q^*\,a\ll 1$ and typically the most important scattering events are those with $q>q^*$.
For one thing, there is more phase space for transitions with higher momentum transfer.
Therefore, interband processes make the main contribution to the self-energy and lead to $m^*<m$ ($|m^*|>|m|$), hence stabilizing the topological phase on a system that was already topological to begin with.
In terms of the bandgap, $E_g^*<E_g$ ($|E_g^*|>|E_g|$).
As $|m|$ becomes larger, the phase space for interband transitions shrinks and intraband contributions begin to dominate, thereby restoring the ``conventional'' behavior of  $|m^*|<|m|$ and $|E_g^*|<|E_g|$.
Thus, $\Sigma_z({\bf 0},0)$ changes sign in the topological phase as a function of $|m|$ (cf. Fig.~\ref{dif_phn}).

In sum, electron-phonon interactions of the deformation potential type favor a topological insulator phase in narrow-gap Dirac insulators with particle-hole symmetry.
The causes behind this unusual effect are the following: (i) the momentum-space texture of the orbital pseudospin changes across a band inversion, (ii) electron-phonon scattering of the deformation potential type conserves spin and pseudospin degrees of freedom, (iii) electron-phonon matrix elements with higher momentum transfer are important (in spite of the larger energy denominators associated to them) due to the increased phase-space for scattering.\cite{caveat}
These three principles might help guide the understanding of how phonons influence the band topology in real Dirac materials with complex band structures.
Moreover, the ideas developed above apply at finite temperature as well.
Given that higher temperature means stronger effective electron-phonon coupling (cf. Eq.~(\ref{geff})), there is the intriguing possibility that heating the system up may drive a trivial insulator into the topological phase.\cite{ion}
Thus far there are no known materials where helical surface states appear only above certain temperature.\cite{dziawa}
In Sec.~\ref{appl} we discuss related phonon-effects which may be accessible in some Dirac materials of current experimental interest.

\begin{figure}
\rotatebox{0}{\includegraphics*[width=\linewidth]{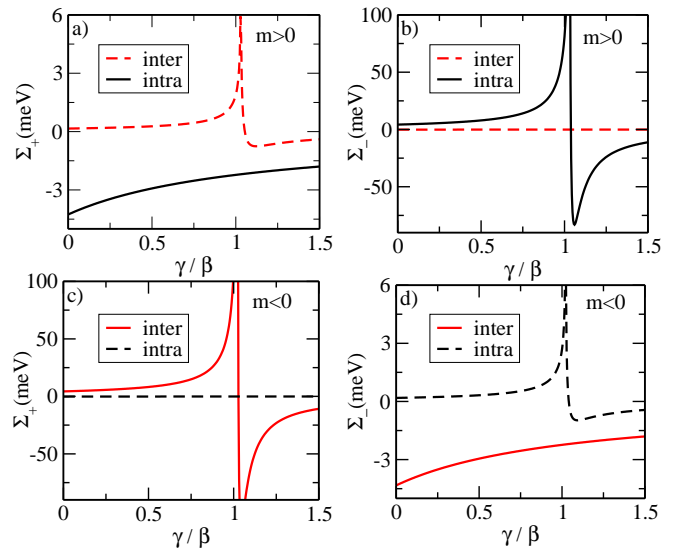}}
\caption{(Color online) Electron-phonon self-energies for the positive- and negative-energy bands of an undoped Dirac insulator, with deformation potential coupling to acoustic phonons.
Intraband and interband contributions are shown separately.
The numerical values of the band and phonon parameters are taken from Ref.~[\onlinecite{ion}].
(a) and (b): Trivial Dirac insulator with $m=10\, {\rm meV}$.
(c) and (d): Topological insulator with $m =-10\,{\rm meV}$.}
\label{se_gamma}
\end{figure}

The preceding arguments relied on particle-hole symmetry and on the absence of itinerant carriers.
Is the mechanism for phonon-induced topological insulation robust under doping ($\epsilon_F\neq 0$) and particle-hole asymmetry ($\gamma\neq 0$)?
On one hand, electron-phonon matrix elements are unchanged by $\epsilon_F\neq 0$ and $\gamma\neq 0$.
Yet, on the other hand, both $\gamma$ and $\epsilon_F$ change the denominators in Eq.~(\ref{intraer}).
In the remaining part of this section, we argue that the conclusions extracted in the preceding paragraph remain robust under moderate particle-hole asymmetry and doping.

Figure~\ref{se_gamma} shows the dependence of $\Sigma_\pm({\bf 0},0)$ on $\gamma$ for an undoped insulator.
When $m>0$, the intraband contribution dominates up to a critical value $\gamma_c\simeq\beta$.
Likewise, when $m<0$, the interband contribution dominates up to $\gamma_c\simeq\beta$.
Therefore, the status quo derived from $\gamma=0$ remains qualitatively unchanged until $\gamma\gtrsim \gamma_c$.
As shown in Fig.~\ref{se_ef}b, the value of $\gamma_c$ may be shifted away from $\beta$ via doping.
When $\gamma=\gamma_c$, the electron-phonon self-energy at zero frequency develops a singularity due to a elastic transition that connects the Fermi-level band eigenstate at ${\bf k}={\bf 0}$ with the band edge at the corner of the Brillouin zone (see also Fig.~\ref{se_ef}a).
Close to the singularity, our perturbative result is unreliable.
For $\gamma\gtrsim\gamma_c$, the system is first an indirect-gap semiconductor and then a semimetal. 
In this regime, the sign of the self-energy is reversed and phonons favor a topologically trivial phase. 
Since $\beta$ is a significant fraction of the electronic bandwidth, we conclude that the mechanism for phonon-induced topological insulation remains robust for moderate particle-hole asymmetry.

Figure~\ref{se_ef}b illustrates the dependence of $\Sigma_z$ on $\epsilon_F$ and $\gamma$.
In the lightly doped materials we are interested in, the screening of the electron-phonon matrix elements by the itinerant carriers is weak.\cite{screen}
Moreover, in this regime the dependence of the self-energy on doping is unremarkable, except close to $\gamma=\gamma_c$.
Hence, once again the simple mechanism discussed above is applicable for moderately doped Dirac systems.
Along the same line, we have confirmed  that the neglect of $\omega_{\bf q}$ on the denominators of Eq.~(\ref{sigma})  is appropriate except near the singularities of the self-energy (not shown).

\begin{figure}
\rotatebox{0}{\includegraphics*[width=\linewidth]{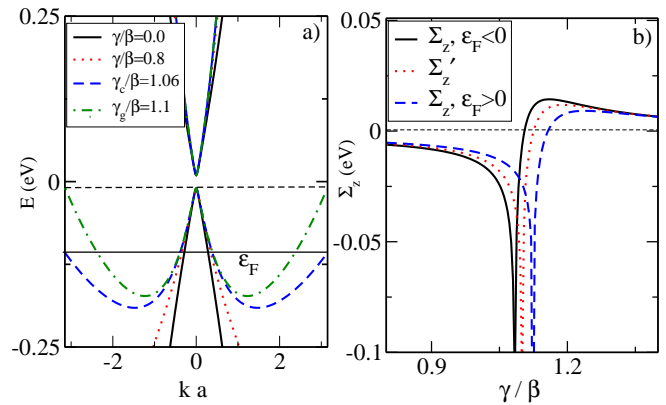}}
\caption{(Color online) (a) Band structure of a toy Dirac insulator (Eq.~(\ref{hD})) as a function of the particle-hole asymmetry parameter $\gamma$.
When $\gamma\simeq\beta$, the phonon-induced renormalization of the bandgap and the Dirac mass develop singularities.
(b) Let us denote the renormalized Dirac mass and bandgap as $m^*=m+\Sigma_z$ and $E_g^*=E_g+\Sigma_z'$, respectively, where $\Sigma_z'$ may be read off Eq.~(\ref{rengap}).
Then, $\Sigma_z$ diverges and changes sign at $\gamma\equiv\gamma_c$, when the band edge at a corner of the Brillouin zone becomes degenerate with the Fermi energy.
Hence, $\gamma_c$ may be tuned by doping.
Likewise, $\Sigma_z'$ diverges and changes sign at $\gamma\equiv\gamma_g$, when the band edge at the zone center becomes degenerate with the band edge at a corner of the Brillouin zone.
Unlike $\gamma_c$, $\gamma_g$ is insensitive to doping.
Such different response of $\gamma_c$ and $\gamma_g$ to doping will play a role in Sec.~\ref{appl}B.}
\label{se_ef}
\end{figure}

\section{Some Applications}
\label{appl}

Thus far we have explained why, in narrow-gap Dirac insulators which are not highly doped or highly particle-hole asymmetric, long-wavelength phonons favor the topological insulating phase.
An experimental signature of this phenomenon would be the emergence of helical surface states {\em above} certain temperature in an insulator that has a topologically trivial ground state.
Traces of this hitherto unobserved phenomenon are more likely to be seen in materials with tunable bandgaps.
HgTe/CdTe quantum wells and BiTl(S$_{1-\delta}$ Se$_\delta$)$_2$ are examples of such materials in two and three dimensions, respectively.
In this section, we discuss signatures of phonon-induced changes in the band topology of these systems.

\subsection{Temperature-dependence of the critical width in HgTe/CdTe quantum wells}
\label{hgte_sec}

Topological insulation in CdTe/HgTe/CdTe quantum wells was predicted\cite{bernevig} by Bernevig, Hughes and  Zhang (BHZ) in 2006. 
The experimental confirmation arrived shortly afterwards.\cite{konig,roth}  
In CdTe, as in most tetrahedral semiconductors,\cite{cardona_book}  the p-type valence band edge ($\Gamma_8$)  lies below the s-type conduction band edge ($\Gamma_6$).
In this ``normal-ordered'' electronic structure, the energy gap is $E_g=1.6\, {\rm eV}$. 
In contrast,  in HgTe, $\Gamma_8$ lies above $\Gamma_6$  and hence the energy gap at the $\Gamma$ point is inverted ($E_g=-0.303\, {\rm eV}$). 
Even though bulk HgTe is semimetallic in absence of strain, it may be coaxed into the insulating phase through quantum confinement in CdTe/HgTe/CdTe quantum wells.
The HgTe layer has a small thickness $d$ along the growth direction $z$ and the heterostructure is translationally invariant in the $xy$ plane.
Accordingly,  ${\bf k}_\perp=(k_x,k_y)$ are good quantum numbers.
At $k_\perp=0$, the lowest-energy subbands in the quantum well are denoted as $E1$ and $H1$, their energies being $E_{\rm E1}$ and $E_{\rm H1}$.
The dispersions of $E1$ and $H1$ with $k_\perp$ are equivalent to those of Dirac fermions with mass $m=(E_{\rm E1}-E_{\rm H1})/2$.  
If $d<d_c$, where $d_c$ is some critical thickness,  the normal ordering of the CdTe electronic structure prevails.
This translates into $m>0$ and trivial insulation.
If $d>d_c$, the ordering between $E1$ and $H1$ subbands is inverted ($m<0$) and the system becomes a topological insulator.
The objective of this subsection is to investigate the effect of electron-phonon interactions on $d_c$.

The natural starting point for such investigation is the BHZ model,\cite{bernevig} which describes the low-energy subbands in the vicinity of $k_\perp=0$.
This model is a special case of Eq.~(\ref{hD}), with the Dirac mass given by $m=(E_{\rm E1}-E_{\rm H1})/2$.
In presence of phonons, $m$ is renormalized to $m^*(T)$ and the critical thickness for the topological transition or crossover is the one for which $m^*(T)=0$.
Anticipating that $|\gamma|<|\beta|$ for all relevant values of $d$ (cf. Fig.~\ref{fig5}b), it follows from Sec.~\ref{phti} that ${\rm Re}[\Sigma_z({\bf 0},i\pi T)] <0$ for any temperature $T$. 
Namely, within the BHZ model, phonons favor the topological phase and therefore one may expect $d_c$ to {\em decrease} as the system is heated. 
This expectation is in stark contrast with the conclusions from a recent theoretical study by Sengupta {\it et al.}, \cite{sengupta}
which has claimed that $d_c$ increases with $T$.
These authors considered the effect of electron-phonon interactions solely through the renormalization of the bandgaps in {\em bulk} HgTe and CdTe.
Such approach is insufficient because it does not capture the influence of phonon-induced transitions between quantum well states.
These transitions, partly included in the electron-phonon self-energy of the BHZ model, are in principle important because they connect states that are close in energy.

We improve on Ref.~[\onlinecite{sengupta}] by taking a two-pronged approach. 
First, we evaluate the temperature-dependence of the band parameters $(m,\alpha,\beta,\gamma)$ appearing in the BHZ model.
This $T-$dependence comes through the phonon-induced renormalization of the bulk CdTe and HgTe bandgaps.
The thickness at which $m(T)$ changes sign will be denoted as $d_c(T)$; this is the quantity that was calculated by Ref.~[\onlinecite{sengupta}]  and found to increase with $T$.
Second, we use the temperature-dependent band parameters as input to calculate the electron-phonon self-energy within the BHZ model.
In this way, we determine the renormalized Dirac mass via $m^*(T)=m(T)+{\rm Re}[\Sigma_z({\bf 0},i\pi T)]$.
The actual critical width $d_c^*(T)$ is defined as the thickness for which  $m^*(T)=0$.
Since ${\rm Re}[\Sigma_z({\bf 0},i\pi T)]<0$, there is the possibility that $d_c^*(T)$ decreases with $T$ even as $d_c(T)$ increases with $T$.

In order to determine the temperature-dependence of the BHZ parameters, we solve the Schr\"{o}dinger equation for a Hg$_{0.32}$Cd$_{0.68}$Te/HgTe/Hg$_{0.32}$Cd$_{0.68}$Te quantum well.
This involves diagonalizing a six-band Kane Hamiltonian\cite{kane} $h({\bf k}_\perp,i\partial_z)$.
The ensuing procedure is identical to that of Ref.~[\onlinecite{bernevig}], except that we take temperature-dependent bandgaps\cite{becker} for bulk HgTe and bulk Hg$_{0.32}$Cd$_{0.68}$Te. 
The solution at $k_\perp=0$ reveals discrete quantum well states, from which $E1\pm$ and  $H1\pm$ have the lowest energies (here $\pm$ labels Kramers partners that are degenerate at $k_\perp=0$ due to time-reversal symmetry). 
The BHZ model follows from applying ${\bf k}\cdot{\bf p}$ perturbation theory around the $k_\perp=0$ solution in the Hilbert space spanned by $\{|E1+\rangle, |H1+\rangle, |E1-\rangle, |H1-\rangle\}$.
By considering only the lowest electronlike and holelike subbands, the BHZ model is two dimensional as far as electrons are concerned.
However, each of the four states forming the low-energy subspace has an associated spinor, whose six components vary with $z$.
We denote these spinors as $\chi_{\sigma\tau}(z)$, where $\sigma=\pm$ and $\tau=E1,H1$.

Figure~\ref{fig5}a displays the calculated dependence of $E_{\rm E1}$ and $E_{\rm H1}$ on $d$ and $T$.
It is apparent that $E_{\rm E1}-E_{\rm H1}$ increases with temperature when $d<d_c$, while $|E_{\rm E1}-E_{\rm H1}|$ decreases with temperature when $d>d_c$.
Hence, the effect of phonons on the {\em bulk} states of CdTe and HgTe favors the normal (i.e. topologically trivial) ordering between $E1$ and $H1$.
Accordingly, Fig. \ref{fig5}c shows that $d_c (T)$ increases with temperature, in agreement  with the result of Sengupta {\it et al.}\cite{sengupta} 
At any rate, $d_c(T)$ is {\em not} the actual critical thickness because we have yet to consider the effect of phonons in the low-energy subspace spanned by $\{|E1\pm\rangle, |H1\pm\rangle\}$.

\begin{figure}
\rotatebox{0}{\includegraphics*[width=\linewidth]{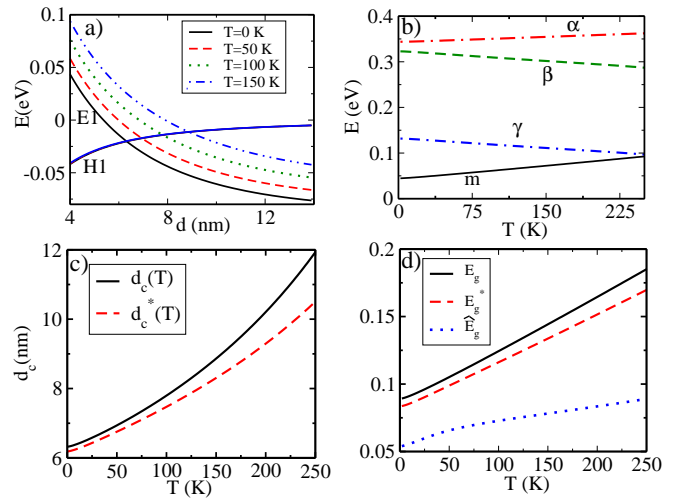}}
\caption{(Color online) (a) Energies of $E1$ and $H1$ subbands ($k_\perp=0$) as a function of 
quantum well width $d$. The parameters needed for the calculation are taken from Ref. \onlinecite{becker}; in addition, we consider $\epsilon_F=0$. 
(b) Temperature-dependence of the band parameters appearing in the BHZ model (Eq.~(\ref{hD})), for $d=4\, {\rm nm}$.
(c) Plot of the critical quantum well (where $m^*(T)$ changes sign) as a function of temperature. 
The solid line ignores phonon-induced intersubband transitions, while the dashed curve partly captures them through the evaluation of the electron-phonon self-energy in the BHZ model.
(d) Energy-difference between $E1$ and $H1$ subbands, as a function of temperature, for a quantum well of thickness $d=4\, {\rm nm}$. 
The solid line is the result without phonon-induced intersubband transitions. 
The dashed line incorporates the phonon effects in the lowest quantum well states.
The doted line ($\hat{E}_g$) takes a slightly larger value of the electron-phonon coupling (by a factor of $\sim 2.5$); the change in the slope of the energy gap as a function of temperature becomes noticeable in this case.
} \label{fig5}
\end{figure}

In order to obtain the actual critical thickness $d_c^*$, we evaluate $\Sigma_z({\bf 0},i\pi T)$ within the BHZ model, using the $T$- and $d$-dependent band parameters derived above.
Owing to a lack of translational invariance along the growth direction, the expression for $\Sigma_z$ differs by a form factor\cite{sarma} from that of Eq.~(\ref{sz}).
This form factor can be derived by recasting the electron density in Eq.~(\ref{rho}) as
\begin{equation}
\label{rho_hgte}
\rho({\bf r})=\frac{1}{A}\sum_{\bf k_\perp q_\perp}\sum_{\tau\tau'\sigma\sigma'}e^{-i{\bf q_\perp}\cdot{\bf r}_\perp} (\chi^*_{\sigma\tau}|\chi_{\sigma'\tau'}) c^\dagger_{{\bf k}_\perp\sigma\tau} c_{{\bf k}_\perp-{\bf q}_\perp\sigma'\tau'},
\end{equation} 
where $A$ is the area of the sample in the $xy$ plane, ${\bf k}_\perp=(k_x,k_y)$, ${\bf q}_\perp=(q_x,q_y)$,  ${\bf r}=({\bf r}_\perp, z)$ and $(\chi^*_{\sigma\tau}|\chi_{\sigma'\tau'})$ is the $z-$dependent scalar product between $\chi_{\sigma\tau}(z)$ and $\chi^*_{\sigma'\tau'}(z)$.
A direct calculation shows that $(\chi^*_{\sigma\tau}|\chi_{\sigma'\tau'})=(\chi^*_{\sigma\tau}|\chi_{\sigma\tau}) \delta_{\tau\tau'} \delta_{\sigma\sigma'}$.
Then, the combination of the first line of Eq.~(\ref{vv}) with Eq.~(\ref{rho_hgte}) yields 
\begin{equation}
\Sigma_z({\bf 0},i\omega)\simeq \sum_{\bf q} F_{q_z} {g_{{\bf q},{\rm eff}}^2\,M_{\bf q_\perp} \over (\epsilon_F+i\omega-d_{0,{\bf q_\perp}})^2-\epsilon_{\bf q_\perp}^2},
\label{eqnhg}
\end{equation}  
where ${\bf q}=({\bf q}_\perp,q_z)$.
The form factor is given by
\begin{equation}
F_{q_z}=\int_{-\infty}^{\infty} dz dz' |\chi(z)|^2 |\chi(z')|^2 e^{-i q_z (z-z')},
\end{equation}
where $|\chi(z)|^2\equiv \sum_{\tau\sigma}(\chi^*_{\sigma\tau}|\chi_{\sigma\tau})$.
In the numerical evaluation of Eq.~(\ref{eqnhg}), we sum over the contributions from three types of electron-phonon interactions: deformation potential coupling to acoustic phonons, deformation potential coupling to optical phonons, and polar optical (Fr\"{o}hlich) coupling.  
Their respective numerical values are listed in App.~\ref{ap_hgte}.

The final outcome of our calculation is collected in Figs.~\ref{fig5}c and \ref{fig5}d, which display $d_c^*$ and the renormalized bandgap as a function of $T$. 
Despite ${\rm Re}[\Sigma_z({\bf 0},i\pi T)]<0$, we find that $d_c^*$ increases with temperature because $m(T)$ increases rather rapidly with $T$ (cf. Fig.~\ref{fig5}d).
That is, for a given quantum well that is topological insulating at $T=0$, increasing temperature produces a crossover into the trivial phase. 
Hence, the conclusion of Ref.~[\onlinecite{sengupta}] is qualitatively correct,  although it overestimates the increase of the critical thickness as a function of temperature. 
In Fig.~\ref{fig5}d we plot the temperature-dependence of the bandgap, $E_{\rm E1}^*-E_{\rm H1}^*$, which is experimentally measurable.\cite{becker} 
In the absence of phonon-induced intersubband transitions,  the gap increases linearly with temperature starting at low temperature. 
However, the inclusion of phonon-induced intersubband transitions and their thermal activation results in a kink in the temperature-dependence of the gap.
The experimental observation of this kink would be an indirect indication of the tendency of phonons to favor a topological phase within the BHZ model.

In sum, the net outcome of electron-phonon interactions in HgTe/CdTe quantum wells is to drive the system closer to the trivial insulating phase.
Had we ignored the temperature-dependence of the band parameters of the BHZ model, we would have wrongly concluded that phonons favor the topological phase.
This is a potentially important lesson that might also impact the theory of topological Anderson insulators,\cite{li} where the effect of disorder on the bulk states of CdTe and HgTe has been overlooked.

We close this subsection with a digression on graphene, which is another canonical two dimensional Dirac insulator.
In graphene, the topological invariant is encoded in the relative sign between the masses of the two Dirac fermions located in the first Brillouin zone.
In inversion and time-reversal symmetric systems, the magnitudes of the two Dirac masses are the same. 
Since electron-phonon interactions do not break any symmetries, they cannot change the sign of one Dirac mass without simultaneously changing the other.
Hence, phonons cannot change the band topology of inversion-symmetric graphene.
In contrast, phonons can alter the band topology of graphene-like systems without inversion symmetry.\cite{li_carbotte}
There is yet another difference between the BHZ model for HgTe/CdTe and the Kane-Mele model\cite{km} for graphene.
In the BHZ model, electron-phonon interactions open a bandgap even when $m=0$ because $M_{\bf k}\simeq m+\beta k^2\neq 0$ for $k\neq 0$.
In graphene, phonons change the bandgap only if the bare gap is nonzero to begin with,\cite{ochoa} because $\beta=0$.\cite{hongki}
Incidentally, these differences are also the reason why strong disorder drives CdTe/HgTe quantum wells into a topological insulating phase,\cite{li} while it drives graphene into a metallic phase.\cite{meyer}

\begin{figure}
\includegraphics[scale=0.3]{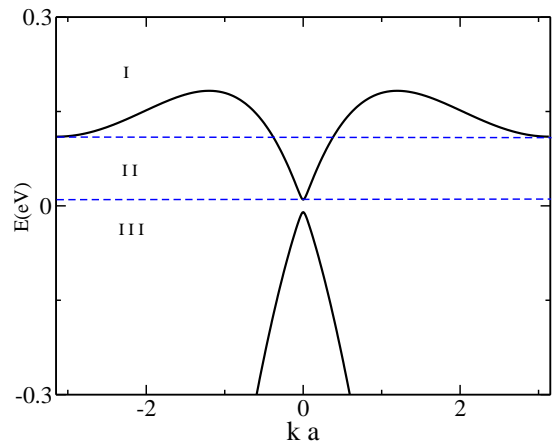}
\caption{(Color online) Qualitative modelling of the low-energy bands\cite{suppl} of  BiTl (S$_{1-\delta}$Se$_{\delta}$)$_2$, using the toy model of Eq.~(\ref{hD}). 
The dashed lines separate three distinct regions of doping, which lead to three different types of phonon-induced effects.
If the Fermi energy is in region I, phonons favor a topologically trivial phase.
If the Fermi energy is in regions II or III, phonons favor a topological phase.
The ``topological proximity effect'' of Ref.~[\onlinecite{hasan1}] takes place when the Fermi energy is in region II.
In contrast, a ``topological antiproximity effect'' is set to occur when the Fermi energy is in region III.} 
\label{fig5a}
\end{figure}

\subsection{Fingerprints of phonons at the topological phase transition of BiTl (S$_{1-\delta}$Se$_{\delta}$)$_2$}
\label{bith}

In 2011, ARPES experiments\cite{hasan} reported evidence for a topological phase transition in BiTl (S$_{1-\delta}$Se$_{\delta}$)$_2$ as a function of the stoichiometric ratio $\delta$.  
The material exhibited helical surface states when $\delta>\delta_c\simeq 0.5$, with the bulk energy gap closing and reopening as $\delta$ was varied.
Due to the finite resolution of ARPES and because of limitations in fine-tuning $\delta$, the value of $\delta_c$ could not be measured accurately.   
However, the authors reinforced the case for $\delta_c\simeq 0.5$  via first-principles electronic structure calculations.
Soon afterwards, the same group completed a more thorough study\cite{hasan1} of $\delta_c$ and announced an unexpected finding: in-gap states emerged at $\delta\simeq0.4-0.5$, {\em prior} to the bulk gap closing. 
Those in-gap states showed no dispersion along the direction normal to the surface and displayed the spin helicity characteristic of topological surface states.
The authors speculated on a ``topological proximity effect'' as a possible origin of the phenomenon.
In this subsection, we argue that it may instead be a fingerprint of electron-phonon interactions.

From the arguments of Sec.~\ref{phti}, we infer that $\delta_c$ (defined as the stoichiometric ratio for which $m^*=0$) must depend on the strength of electron-phonon interactions.
Moreover, first-principles electronic structure calculations\cite{suppl} show that the conduction band at the $X$ point of the bulk Brillouin zone is nearly degenerate with the conduction band at the $\Gamma$ point, which in turn suggests that the phonon-induced renormalization of the bandgap can be significant. 

Empirically, there are two ways to verify that $\delta_c$ depends on electron-phonon interactions.
On one hand, the measured value of $\delta_c$ (which inevitably incorporates phonon effects) should be different from the value predicted by existing {\em ab-initio} calculations\cite{suppl} (which ignore phonons).
Admittedly, electron-phonon interactions are not the only agents that can shift the value of $\delta_c$ with respect to the non-interacting case: Coulomb interactions and short-ranged non-magnetic disorder may have an impact as well.
On the other hand, if phonons are at play, the measured $\delta_c$ should be strongly temperature-dependent on the scale of the Debye temperature.
The observation of such temperature-dependence would in fact be a true smoking gun for phonon-induced effects in the band topology of BiTl (S$_{1-\delta}$Se$_{\delta}$)$_2$, because neither static disorder nor Coulomb-like electron-electron interactions should produce a significantly temperature-dependent effect.
Arguably, it is not easy to measure $\delta_c (T)$ because the thermal smearing of the quasiparticle bands prevents locating the exact point where surface states emerge.
Nevertheless, it should be relatively easy to measure the temperature-dependence of the bulk bandgap when $\delta$ is sufficiently far from $\delta_c$.
A phonon-induced reduction of $\delta_c$ would manifest itself through $d |E_g^*|/dT<0$  (if $\delta\ll \delta_c$) and $d |E_g^*|/dT>0$ (if $\delta\gg \delta_c$).
Instead if phonons increased $\delta_c$ with respect to the non-interacting case, the observed temperature-dependence of the bandgap would be of opposite sign. 

We model the low-energy electronic structure of BiTl (S$_{1-\delta}$Se$_{\delta}$)$_2$ qualitatively by a lattice version of Eq.~(\ref{hD}).
Upon selecting a value of $|\gamma|\lesssim|\beta|$, we can partially mimic the realistic scenario where the conduction band at the Brillouin zone edge is nearly degenerate with that of the zone center.
The corresponding band structure is shown in Fig.~\ref{fig5a}. 
Therein, we have identified three different regions.
When the Fermi energy is in region I,  $|\gamma_c|<|\gamma|<|\gamma_g|$ and thus $\Sigma_z>0>\Sigma_z'$.
For the definitions of $\gamma_c$, $\gamma_g$ and $\Sigma_z'$, see the caption of Fig.~\ref{se_ef}.
In this case, phonons favor a trivial insulating phase. 
In contrast, in both regions II and III phonons favor a topological insulating phase.
When the Fermi energy is in region II,  $|\gamma|<\gamma_c|<|\gamma_g|$ and  $0>\Sigma_z'>\Sigma_z$.
Since $|\Sigma_z|>|\Sigma_z'|$, the Dirac mass renormalizes more strongly than the bulk energy gap.
When the Fermi energy is in region III,  $|\gamma|<|\gamma_g|<|\gamma_c|$ and $0>\Sigma_z>\Sigma_z'$.
Here, the Dirac mass renormalizes less strongly than the bulk energy gap.  
The consequences of this will be discussed below.

Figure~\ref{bitl}(c) illustrates the energies of the conduction and valence band edges (cf. Eq.~(\ref{rengap})) as a function of $\delta$, both in presence and absence of electron-phonon interactions.
Clearly, phonons favor the topological insulating phase. 
Along the same line, Fig.~\ref{bitl}(d) shows the reduction of $\delta_c(T)$ as a function of temperature and predicts the emergence of helical surface states beyond a crossover temperature, when $\delta<0.5$. 

Throughout these plots, we have extrapolated the experimental data of Ref.~[\onlinecite{hasan}] into a linear relation between the bare Dirac mass $m$ (which is half the non-interacting bandgap) and $\delta$.
More complicated $m(\delta)$ functions would not change our conclusions qualitatively.
In addition, we have neglected the thermal expansion of the lattice.
Thermal expansion renders all the band parameters of Eq.~(\ref{hD}) temperature-dependent even before the inclusion of the electron-phonon self-energy.
In the previous subsection, we have learned that such ``extraneous'' temperature-dependence can potentially revert the trend that one would have anticipated solely from the calculation of the self-energy.
We justify our approach on the basis that, in most semiconductors, the contribution of thermal expansion to bandgap renormalization is small enough that its neglect poses no risk for qualitative error.\cite{exception}

\begin{figure}
\rotatebox{0}{\includegraphics*[width=\linewidth]{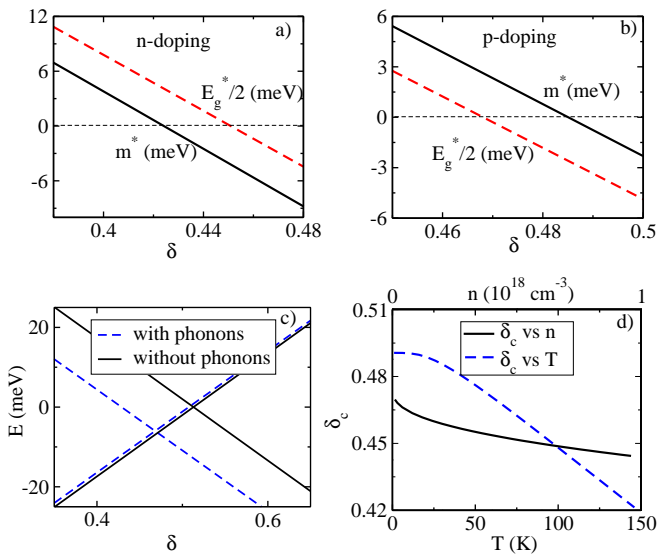}}
\caption{(Color online) 
(a) and (b): Zero-temperature renormalized Dirac mass $m^*$ and half-bandgap $E_g^*/2$ of BiTl (S$_{1-\delta}$Se$_{\delta}$)$_2$ as a function of $\delta$, when the Fermi energy is (a) in region II  and (b) in region III.
The electron- and hole densities in (a) and (b) are $n=10^{19} {\rm cm}^{-3}$ and $p=10^{19} {\rm cm}^{-3}$, respectively.
In (a), the emergence of topological surface states precedes the band inversion (``topological proximity effect'').
In (b), the emergence of topological surface states succeeds the band inversion (``topological antiproximity effect'').
(c) Influence of electron-phonon interactions on the renormalized conduction and valence band edges.
The solid (dashed) lines correspond to the absence (presence) of electron-phonon interactions. 
Phonons decrease the magnitude of the bandgap in the trivial phase, while they enhance it in the topological phase.
The mechanism behind this effect has been explained in Sec.~\ref{phti}. 
(d) Temperature- and density-dependence of the critical stoichiometric ratio. 
Throughout this figure we have taken $\gamma=-0.22 {\rm eV}$; all other band and phonon parameters are the same as in Ref.~[\onlinecite{ion}].}
 \label{bitl}
\end{figure}

We now propose a possible explanation for the ``topological proximity effect'' reported in Ref.~[\onlinecite{hasan1}].
Topological surface states take place when $m^*<0$, while the bulk gap measured in ARPES corresponds to $|E_g^*|$.  
As mentioned in Sec.~\ref{dirac}, $m^*\neq E_g^*/2$ in presence of interactions.
In Figs.~(\ref{bitl})a and (\ref{bitl})b, we plot $m^*$ and $E_g^*$ as a function of $\delta$ for different electron densities.
Since S and Se belong to the same group in the periodic table, changing $\delta$ does not change the carrier density.
We find that, when the system is neutral or hole-doped (cf. Fig.~\ref{bitl}b), i.e. when the Fermi energy is located in region III, there exists an interval of $\delta$ for which $E_g^*<0$ and $m^*>0$.
This is a consequence of $0<\Sigma_z<\Sigma_z'$ and it indicates an onset of topological surface states after (and not simultaneously with) the occurrence of a band inversion.
In contrast, in a weakly electron-doped system (such that the Fermi energy is in region II, cf. Fig.~\ref{bitl}a), we find $m^*<0$ and  $E_g^*>0$ for $\delta\in(\delta_c,\delta_c+\Delta\delta)$.
Assuming a reasonable strength of the electron-phonon coupling,\cite{ion} $\Delta\delta\simeq 0.2$, the bandgap at $\delta=\delta_c$ is $E_g^*\simeq 10\,{\rm meV}$ and
the band broadening due to electron-phonon interactions remains small.
In such scenario, the onset of topological surface states {\em precedes} the occurrence of a band inversion.
These results appear to be consistent with Ref.~[\onlinecite{hasan1}], whose samples are electron-doped.
Coulomb interactions are unlikely to be responsible for this effect because their associated self-energy has a weak frequency-dependence.
On the other hand, although the disorder self-energy is frequency-dependent, its temperature-dependence is negligible.
Hence, the observation of a temperature- and density-dependent $\Delta\delta$ would confirm the key role of phonons.

Finally, we comment on the imaginary part of the self-energy, which gives the broadening of the renormalized quasiparticle bands.
In order to observe phonon-induced topological surface states, it is essential that the phonon-induced broadening of the bandgap at $k\simeq 0$ be small. 
It is easy to show that, so long as the width of region II in Fig.~\ref{fig5a} is larger than the characteristic phonon energy scale, the phonon-induced band broadening of the $k=0$ bandgap is negligible.
This is the case for the parameter values taken in Fig.~\ref{bitl}. 

\section{Summary and discussion}
\label{summ}

Electron-phonon interactions can induce topological insulation in a narrow-gap Dirac material with an intrinsically trivial electronic structure.
The essential ingredients behind this phenomenon are the following: (i) a direct and small bandgap, (ii) the change in the momentum-space texture of the band eigenstates from the trivial to the topological phase, (iii) the conservation of spin and orbital degrees of freedom in electron-phonon scattering processes involving long wavelength phonons, (iv) the importance of electron-phonon matrix elements with high momentum transfer (due to the increased scattering phase-space associated to them.)

Together, these ingredients produce a peculiar outcome: while the leading electron-phonon matrix elements in a trivial Dirac insulator are those in which a phonon scatters an electron within the same band, the dominant matrix elements in a narrow-gap topological insulator are those in which a phonon scatters an electron between the conduction and the valence band.
A direct consequence of this peculiarity in the electron-phonon matrix elements is that the bandgap of a trivial (topological) Dirac insulator decreases (increases) as temperature is raised.
This, in turn, anticipates the emergence of helical surface states beyond a crossover temperature in a Dirac insulator with a topologically trivial ground state.

The above mechanism is not exclusive to phonons and can be transferred to spin-independent disorder and Coulomb interactions.
The large dielectric constant of common Dirac insulators implies that the effect of Coulomb interactions in band topology will often be small in comparison to that of electron-phonon interactions.
As for disorder, the signatures of topological Anderson insulation must be accessed by transport experiments because it is difficult to measure the bandgap as a function of random impurity concentration.
Such transport experiments are rather contrived due to the unintended bulk doping that is prevalent in many Dirac materials. 
In contrast, measuring the temperature-dependence of a bulk bandgap is routine.
In sum, the main aspect that sets phonons apart from other agents is that their influence on band topology is significantly temperature-dependent.

Signatures of phonon-induced topological insulation have not yet been confirmed in experiment.
What are the materials to look for and what should be measured?
In principle, any Dirac insulator with a direct and small bandgap can display the phenomenon.
A small gap is necessary because otherwise intraband electron-phonon scattering processes dominate and lead to a decrease of the bandgap regardless of the band topology.
Since the typical electron-phonon self-energies are $\sim 10-100\, {\rm meV}$, the most spectacular phonon effects occur in the vicinity of a topological phase transition.
As of this writing, there is an increased number of Dirac materials where experimentalists are able to apply pressure or change the stoichiometry continuously in order to tune the bandgap from the trivial to the topological phase, and vice versa.
Should the bandgap decrease (increase) with raising temperature in the trivial (topological) side of the transition, this would confirm that phonons favor a toplogical insulating phase.
Recent experiments in BiTl(S$_{1-\delta}$Se$_\delta$) have reported that topological surface states emerge at $\delta=\delta_c-\Delta\delta <\delta_c$, whereas the bulk gap closes at $\delta=\delta_c$.  
Does $\Delta\delta$ depend on carrier concentration and on temperature?
As explained in Sec.~\ref{appl}B, an affirmative response would suggest that phonons are behind the observed effect.

We conclude by assessing the limitations of our theory, which relies on three assumptions: (i) the lowest conduction band and the highest valence band are well separated in energy from the rest of the bands, (ii) short wavelength phonons can be neglected, (iii) the leading phonon effects originate from the coupling between the density of electrons and the lattice deformation.

The first assumption, which enables the use of low-energy effective models to arrive at a simple picture of phonon-induced topological insulation, is often justified
in the vicinity of a topological phase transition.
When phonon-induced transitions to high-energy bands become significant, we anticipate that a formally similar low-energy effective model will still be applicable, albeit with renormalized band parameters.
Should this renormalization be strong enough, the influence of phonons on band topology might be reversed.
For instance, in CdTe/HgTe quantum wells (cf. Sec.~\ref{appl}A), the net effect after including electron-phonon interactions in high energy bands is that phonons favor the {\em trivial} phase.

The neglect of short wavelength phonons (and hence of the Debye-Waller term) has been done on the basis of simplicity and is nearly universal in textbooks.
Yet, in real materials, the Debye-Waller contribution to bandgap renormalization may be of the same order as the self-energy contribution discussed in this work.
Recognizing that Debye-Waller processes involve vertical (zero-momentum-transfer) interband transitions,\cite{cohen} which should be insensitive to the occurrence of a band inversion, we speculate that the Debye-Waller term will renormalize the bandgap but not the Dirac mass (thereby not affecting the band topology).

Finally, the third assumption above ignores phonon-induced processes that alter electronic hopping amplitudes. 
Such terms have been discussed in graphene\cite{ando} and can also exist in Dirac insulators.
They can be incorporated into our theory in an {\em ad-hoc} fashion by promoting the electron-phonon coupling from an identity matrix in spin and orbital space to a matrix with off-diagonal elements.
The off-diagonal elements might lead to phonon-induced {\em trivial} insulation.
For instance, if a phonon mode exists which flips the orbital pseudospin, then this mode will directly oppose phonon-induced topological insulation.
However, it is likely that the off-diagonal matrix elements of the electron-phonon coupling in the spin and orbital space are often small compared to the diagonal elements, much like in graphene.

In the future, it would be desirable to resort to first-principles calculations that relax our assumptions.
Some initial efforts along this direction are underway.\cite{jkim}
It is our hope that the basic insights for phonon-induced topological insulation, unearthed here in the context of a toy model, will remain relevant after considering the full complexity of the electronic and phononic band structures.

\acknowledgements
We acknowledge useful discussions or correspondence  with S. Adam, G. Antonius, A. Bansil, M. C${\hat {\rm o}}$t\'e, D. Hsieh, Z. Li, H. Lin and R. Nourafkan.
The numerical calculations were performed on computers provided by Calcul Qu\'ebec and Compute Canada.
This research has been financially supported by Universit\'e de Sherbrooke, RQMP and Canada's NSERC.

\appendix
\section{Debye-Waller contribution}
\label{ap_dw}

Let us expand the electron-ion potential with respect to small displacements of the ions from their equilibrium positions:
\begin{align}
\label{ei}
&V_{ei}({\bf r}-{\bf R}_j^{(0)}-{\bf Q}_j) = V_{ei}({\bf r}-{\bf R}^{(0)}_j)\nonumber\\
&-{\bf Q}_j\cdot{\boldsymbol\nabla} V_{ei}({\bf r}-{\bf R}_j^{(0)})+\frac{1}{2}({\bf Q}_j\cdot{\boldsymbol\nabla})^2 V_{ei}({\bf r}-{\bf R}_j^{(0)})+...
\end{align}
In textbooks,\cite{mahan} only the first order term in the expansion of Eq.~(\ref{ei}) is kept, which leads to ${\cal V}^{(1)}$ (cf. Eq.~(\ref{vv})) in the limit of long-wavelength phonons, with
\begin{equation}
g_{\bf q}=\sqrt{\frac{\hbar}{2 \rho_A V \omega_{\bf q}}} {\bf q}\cdot{\bf e}_{\bf q} V_{ei}({\bf q})
\end{equation}
Thereafter, the effect of electron-phonon interactions on physical observables is computed through second order perturbation theory in ${\cal V}^{(1)}$.
The term of order $Q^2$ in Eq.~(\ref{ei}) is ignored in textbooks.
However, if treated in first order perturbation theory, it contributes at the same order as the perturbation kept in textbooks.
This contribution receives the name of ``Debye-Waller term''.
In this Appendix, we show that the Debye-Waller term vanishes in the limit of long-wavelength phonons.

Following the same steps\cite{mahan} as in the derivation of ${\cal V}^{(1)}$, the electron-phonon interaction emerging from the $O(Q^2)$ term in Eq.~(\ref{ei}) can be written as
\begin{widetext}
\begin{align}
\label{v2}
{\cal V}^{(2)}=\frac{1}{2}\sum_{{\bf k},{\bf q},{\bf G}}\rho_{{\bf q}+{\bf G}} V_{ei}({\bf q}+{\bf G}) [({\bf q}+{\bf G})\cdot{\bf e}_{\bf k}] [({\bf q}+{\bf G})\cdot{\bf e}_{\bf k-q}]
\sqrt{\frac{\hbar}{2\rho_A V \omega_{\bf k}}}\sqrt{\frac{\hbar}{2\rho_A V \omega_{-\bf k+q}}} (a_{\bf k}+a^\dagger_{-\bf k})(a_{-{\bf k}+{\bf q}}+a^\dagger_{{\bf k-q}}),
\end{align}
where ${\bf G}$ is a reciprocal lattice vector.
If the short-wavelength phonons are neglected (which implies keeping only the $G=0$ term\cite{mahan} in the sum of Eq.~(\ref{v2})), then Eq.~(\ref{v2}) reduces to the second line of Eq.~(\ref{vv}), with
\begin{equation}
\lambda_{{\bf k}{\bf q}}=\sqrt{\frac{\hbar}{2\rho_A V \omega_{\bf k}}}\sqrt{\frac{\hbar}{2\rho_A V \omega_{-\bf k+q}}}V_{ei}({\bf q}) ({\bf q}\cdot{\bf e}_{\bf k})\, ({\bf q}\cdot{\bf e}_{\bf k-q}).
\end{equation}
The change in the electronic structure due to ${\cal V}^{(2)}$ involves the following expectation value of the phonon operators
\begin{equation}
\langle  (a_{\bf k}+a^\dagger_{-\bf k})(a_{-{\bf k}+{\bf q}}+a^\dagger_{{\bf k-q}})\rangle\nonumber\\
 =  \langle a_{\bf k} a^\dagger_{{\bf k-q}} + a^\dagger_{-\bf k} a_{-{\bf k}+{\bf q}} \rangle \propto \delta_{{\bf q},{\bf 0}}.
\end{equation}
Since $\lambda_{{\bf k}{\bf q}}=0$ for $q=0$, it follows that ${\cal V}^{(2)}$ makes a vanishing contribution to the renormalized electronic energy levels.
This conclusion does not hold when $G\neq 0$ terms are kept in Eq.~(\ref{v2}).

For completeness, we show an alternative derivation for the vanishing of the Debye-Waller term.
Following the work of Allen and Cardona in Ref.~[\onlinecite{cohen}], the Debye-Waller term involves a vertical interband matrix element of the type
\begin{equation}
\langle \psi_{{\bf k} n}|\frac{\partial V_{ei}({\bf r}-{\bf R})}{\partial{\bf R}}|\psi_{{\bf k} n'}\rangle
=i\sum_{{\bf k'}} {\bf k}' V_{ei}({\bf k'}) e^{-i {\bf k}'\cdot{\bf R}}\int_{\rm all} d{\bf r}\, \psi_{{\bf k}n}^*({\bf r}) \psi_{{\bf k}n'}({\bf r}) e^{i {\bf k}'\cdot{\bf r}}\nonumber
=i\sum_{\bf G} {\bf G}\, V_{ei}({\bf G})\int_{\rm cell} d{\bf r}\, e^{i {\bf G}\cdot{\bf r}} u_{{\bf k}n}^*({\bf r}) u_{{\bf k} n'}({\bf r}),
\end{equation}
where ${\bf R}$ is the ion coordinate and the spatial integrals in first and second equalities are over the entire crystal and over the unit cell, respectively.
In addition, we have used the Bloch's theorem: $\psi_{{\bf k}n}({\bf r})=\exp(i {\bf k}\cdot{\bf r}) u_{{\bf k}n}({\bf r})$. 
If the eigenstates are plane waves, the spatial integration selects $G=0$ and the matrix element vanishes.
Even for non-plane-wave eigenstates, the $G=0$ term is clearly zero.
Neglecting $G\neq 0$ terms (and hence ignoring the Debye-Waller term) is justified if the electron-ion potential changes slowly in space (which is the case in the deformation potential approximation) and/or if the Bloch eigenstates are approximately plane waves (which is an acceptable approximation for some simple metals, though not for most semiconductors). 

\end{widetext}

\section{Electron-phonon coupling parameters for HgTe/CdTe quantum wells}
\label{ap_hgte}

For reference, in this Appendix we list the various types of $g_{\bf q}$ that were used in Sec.~\ref{appl}A.
Since we are interested in long wavelength phonons and since the quantum well states have no degeneracies beyond the standard spin degeneracy, we consider the coupling to longitudinal modes only.
Because crystals with zincblend structure contain two atoms per unit cell, we consider one longitudinal acoustic phonon and one longitudinal optical phonon.
The optical phonon can couple to electrons through deformation potential and through Fr\"ohlich-type interaction.
The numerical parameters are quoted from Ref.~[\onlinecite{lhuillier}].
 
In the case of deformation potential coupling to longitudinal acoustic phonons, 
\begin{equation}
g_{\bf q}=\sqrt{\frac{\hbar C_{\rm ac}^2 q^2}{2\rho_A V \omega_{\bf q}}},
\label{eqnac}
\end{equation}
where $\rho_A=8100 {\rm kg/m}^3$ is the atomic mass density, $C_{\rm ac}=5\,{\rm eV}$ is the acoustic deformation potential coupling constant,  $\omega_{\bf q}=c_s q$ is the phonon frequency and $c_s=2100\,{\rm m/s}$ is the sound velocity.

In the case of deformation potential coupling to non-polar optical phonons, 
\begin{equation}
 g_{\bf q}=\sqrt{\frac{\hbar C_{\rm op}^2/a^2}{2\rho_A V \omega_0}},
\end{equation}
where $C_{\rm op}=20\,{\rm eV}$ is the optical deformation potential, $\omega_0=17\,{\rm meV}$ is the optical phonon frequency and $a\simeq 0.646\, {\rm nm}$ is the lattice constant for HgTe.  
In the case of the Fr\"ohlich coupling to polar optical phonons, 
\begin{equation}
g_{\bf q}=\sqrt{\frac{e^2\hbar \omega_0}{ V q^2}\left({1 \over \epsilon_{\infty}}-{1\over \epsilon_{0}}\right)},
\end{equation}
where $\epsilon_{\infty}=14$ and $\epsilon_{0}=20$ are high-frequency and static dielectric constants (in units of vacuum permittivity).




\begin{thebibliography}{99}
\bibitem{scba} A calculation based on the self-consistent Born approximation confirms this conclusion. At any rate, the underlying assumption is that there are no lattice instabilities as the renormalized electronic gap closes; in two and three dimensions, such assumption is justifiable.
\bibitem{caveat} By ``higher-momentum transfer'', we mean $q\gg \sqrt{|m|/\beta}$. For Dirac insulators near a topological phase transition, this condition is satisfied even when momentum transfer is small compared to a reciprocal lattice vector.  Accordingly, our approximate long-wavelength expressions for the electron-phonon interaction (Eq.~(\ref{vv})) remain applicable.
\end{thebibliography}
\end{document}